\documentclass[twoside,twocolumn,9pt]{article}
\usepackage{extsizes}
\usepackage[super,sort&compress,comma]{natbib} 
\usepackage[version=3]{mhchem}
\usepackage[left=1.5cm, right=1.5cm, top=1.785cm, bottom=2.0cm]{geometry}
\usepackage{balance}
\usepackage{widetext}
\usepackage{times,mathptmx}
\usepackage{sectsty}
\usepackage{graphicx} 
\usepackage{lastpage}
\usepackage[format=plain,justification=raggedright,singlelinecheck=false,font={stretch=1.125,small,sf},labelfont=bf,labelsep=space]{caption}
\usepackage{float}
\usepackage{fancyhdr}
\usepackage{fnpos}
\usepackage[english]{babel}
\usepackage{array}
\usepackage{droidsans}
\usepackage{charter}
\usepackage[T1]{fontenc}
\usepackage[usenames,dvipsnames]{xcolor}
\usepackage{setspace}
\usepackage[compact]{titlesec}

\usepackage{epstopdf}

\definecolor{cream}{RGB}{222,217,201}

\begin{document}

\pagestyle{fancy}
\thispagestyle{plain}
\fancypagestyle{plain}{

\renewcommand{\headrulewidth}{0pt}
}

\makeFNbottom
\makeatletter
\renewcommand\LARGE{\@setfontsize\LARGE{15pt}{17}}
\renewcommand\Large{\@setfontsize\Large{12pt}{14}}
\renewcommand\large{\@setfontsize\large{10pt}{12}}
\renewcommand\footnotesize{\@setfontsize\footnotesize{7pt}{10}}
\makeatother

\renewcommand{\thefootnote}{\fnsymbol{footnote}}
\renewcommand\footnoterule{\vspace*{1pt}%
\color{cream}\hrule width 3.5in height 0.4pt \color{black}\vspace*{5pt}} 
\setcounter{secnumdepth}{5}

\makeatletter 
\renewcommand\@biblabel[1]{#1}            
\renewcommand\@makefntext[1]%
{\noindent\makebox[0pt][r]{\@thefnmark\,}#1}
\makeatother 
\renewcommand{\figurename}{\small{Fig.}~}
\sectionfont{\sffamily\Large}
\subsectionfont{\normalsize}
\subsubsectionfont{\bf}
\setstretch{1.125} 
\setlength{\skip\footins}{0.8cm}
\setlength{\footnotesep}{0.25cm}
\setlength{\jot}{10pt}
\titlespacing*{\section}{0pt}{4pt}{4pt}
\titlespacing*{\subsection}{0pt}{15pt}{1pt}

\fancyfoot{}
\fancyfoot[RO]{\footnotesize{\sffamily{1--\pageref{LastPage} ~\textbar  \hspace{2pt}\thepage}}}
\fancyfoot[LE]{\footnotesize{\sffamily{\thepage~\textbar\hspace{3.45cm} 1--\pageref{LastPage}}}}
\fancyhead{}
\renewcommand{\headrulewidth}{0pt} 
\renewcommand{\footrulewidth}{0pt}
\setlength{\arrayrulewidth}{1pt}
\setlength{\columnsep}{6.5mm}
\setlength\bibsep{1pt}

\makeatletter 
\newlength{\figrulesep} 
\setlength{\figrulesep}{0.5\textfloatsep} 

\newcommand{\topfigrule}{\vspace*{-1pt}%
\noindent{\color{cream}\rule[-\figrulesep]{\columnwidth}{1.5pt}} }

\newcommand{\botfigrule}{\vspace*{-2pt}%
\noindent{\color{cream}\rule[\figrulesep]{\columnwidth}{1.5pt}} }

\newcommand{\dblfigrule}{\vspace*{-1pt}%
\noindent{\color{cream}\rule[-\figrulesep]{\textwidth}{1.5pt}} }

\makeatother


\twocolumn[
  \begin{@twocolumnfalse}
\sffamily
\begin{tabular}{m{0.01cm} p{17.99cm} }

& \noindent\LARGE{\textbf{Phase separation in binary mixtures of active and passive particles}} \\
\vspace{0.3cm} & \vspace{0.3cm} \\

 & \noindent\large{Pritha Dolai,\textit{$^{a,b}$} Aditi Simha,\textit{$^{b}$} and Shradha Mishra\textit{$^{c}$}} \\
\vspace{2.5cm}  & \vspace{0.3cm} \\
& \noindent\normalsize{We study binary mixtures of small active and big passive athermal particles interacting via soft repulsive forces 
on a frictional substrate. Athermal self propelled particles are known to phase separate into a dense aggregate and a dilute gas-like phase at fairly low
packing fractions. Known as {\emph {motility induced phase separation}}, this phenomenon
governs the behaviour of binary mixtures for small to intermediate size ratios of
the particle species. 
An effective attraction between passive particles, due to the surrounding active medium, leads to true phase separation for large size ratios and volume fractions of active particles.
The effective interaction between active and passive particles can be attractive or repulsive at short range depending on the size ratio and volume 
fractions of the particles. This affects the clustering of passive particles. We find three distinct phases based on the spatial distribution of passive particles. 
 The cluster size distribution of passive particles decays exponentially in the {\emph{homogeneous phase}}.
It decays as a power law with an exponential cutoff in the {\emph{clustered phase}} and tends to a power law as the system approaches the transition to the 
{\emph{phase separated state}}.
We present a phase diagram in the plane defined by the
size ratio and volume fraction of passive particles.}

\end{tabular}

 \end{@twocolumnfalse} \vspace{0.6cm}

  ]

\renewcommand*\rmdefault{bch}\normalfont\upshape
\rmfamily
\section*{}
\vspace{-1cm}


\footnotetext{\textit{$^{a}$International Centre for Theoretical Sciences, Hesaraghatta Hobli, Bengaluru 560089, India; E-mail: pritha.dolai@icts.res.in }}
\footnotetext{\textit{$^{b}$  Department of Physics, Indian Institute of Technology Madras, Chennai 600036, India; E-mail: phyadt@iitm.ac.in}}
\footnotetext{\textit{$^{c}$ Department of Physics, Indian Institute of Technology BHU, Varanasi-221005, India; E-mail: smishra.phy@itbhu.ac.in }}






\section{Introduction }\label{intro}

Active systems or systems consisting of self propelled particles have been a
subject of great interest and research in recent years. Examples of such systems range from very small scales, a few microns in size,
like molecular motors, bacterial colonies to very large scales upto a few kilometers like bird flocks, human crowds, animal herds and such others. 
Interaction among different agents in these systems is mainly short range steric repulsion. In the presence of activity  they show
interesting collective behaviour like coherent motion, phase separation without any external drive and such like\cite{Simha,Sriram1}. 
It was shown recently that athermal self-propelled particles interacting solely via steric repulsion,  
phase separate into a dense solid-like phase and a dilute gas phase at very low volume fractions, much 
before the system reaches close packing ~\cite{Fily1,Fily2}. This phenomenon, unique to active systems, has come to be known as
 {\emph {motility induced phase separation}} (MIPS)
~\cite{Fily1,Redner,Cates1,Tailleur1,Gonnella1,Gonnella2,Berthier,Cates2}\, and been observed in a number of 
simulations of self-propelled particles ~\cite{Fily2,Speck1,Peruani2} and also 
been realised experimentally in a system of synthetic colloids ~\cite{Speck2,Palacci1}\,. 

Active particles have also been found to induce effective interactions between 
large passive colloidal particles, similar to the depletion interaction in equilibrium mixtures. Known as {\emph {active depletion}}, 
the range, strength, and sign of these interactions are crucially dependent on the shape and size of the colloidal particle. 
Colloidal rods experience a long-ranged predominantly attractive interaction while colloidal disks feel a repulsive force that is 
short-ranged in nature and grows in strength with the size ratio of the colloids and active particles ~\cite{Angelani,Ni,Harder}\,.

Phase separation in equilibrium binary mixtures is known to occur even when 
the interactions are purely repulsive \cite{Frenkel,biben}. 
This has been attributed to an attractive depletion force \cite{Asakura1,Asakura} between large particles in the mixture which is due 
to the unbalanced osmotic pressure exerted on them by the surrounding small particles. Phase separation in these mixtures is an entropy-driven 
first order transition and occurs for size 
ratios greater than 5 in binary mixtures of hard spheres. \cite{Frenkel,biben}\,. 

These two phenomena, depletion induced binary phase separation and motility
induced phase separation (MIPS), motivate us to choose as our system a binary 
mixture of active and passive discs. Experimentally, passive particles in active systems have been used as tracers to probe 
and quantify the dynamics of active particles in active suspensions. 
The motion of micron-sized passive beads suspended in an active bath was first studied by 
Wu ${\textit {et al.}}$ ~\cite{Libchaber}\,. 
The effective diffusivities of passive particles was found to be larger than their thermal diffusivities. 
More recently Patteson ${\textit {et al.}}$ have experimentally shown that, unlike classical results, the 
diffusivities of passive particles in an active bath is non-monotonic in particle size ~\cite{Gopinath}\,. 
Activity induced phase separation has also been observed in Brownian dynamics simulations of monodisperse mixtures of active and 
passive particles ~\cite{Cates3,Bechinger,Simon}\,. Phase segregation of passive advective particles has been observed in a medium 
of active polar actomyosin filaments ~\cite{Rao}\,.

In this paper, we study the  phase behaviour of athermal mixtures of active and passive particles. 
All particles in our system interact via soft repulsive forces. Active particles are self driven and have a 
self-propulsion velocity $\vec{v}$. Their speed $v$ remains constant whereas their direction 
changes randomly over a time scale $\tau_r=\nu_{r}^{-1}$, the inverse of the rotational diffusion constant. 
Neither species is subjected to random translational 
noise and is hence athermal. In addition to this, the two species of particles differ in size (or radius).
The mixture is analyzed for various size ratios and compositions. 
Although active-passive mixtures have been studied earlier, most of them are limited to non-spherical  passive particles and when they are few in number 
~\cite{Ni,Galajda,Kaiser}. 
A full phase diagram of the mixture in terms of its defining parameters is still lacking. What happens when the passive particles are comparable in number or volume fraction to
the active particles? In our present study,  
we address the following questions. What is the effective interaction between passive particles in the mixture ~\cite{Naji}\, and how does it vary with the system parameters? 
What is the nature of interaction between an active and passive particle? How does motility induced phase separation affect this? What are the various phases exhibited by the 
system in the parameter space of ($\phi_a,\, \phi_p,\, s\,$), the volume fractions of active and passive particles respectively and the size ratio of particles in the mixture? 
Does the mixture phase separate as in equilibrium binary mixtures?

The tendency of active particles to undergo motility induced phase separation
strongly influences the phase behaviour of our binary mixture.
An effective attraction between big passive particles leads to true phase separation  for large size ratios and volume fractions of active particles.
The effective interaction between active and passive particles can be attractive or repulsive at short range depending on the size ratio and volume 
fractions of the particles. This affects the clustering of passive particles. In the former, the cluster size decreases with increasing size ratio while in the latter, 
it increases with increasing size ratio. We find three distinct phases based on the spatial distribution of passive particles. 
 The cluster size distribution (CSD) of passive particles decays exponentially in the {\emph{homogeneous mixed phase}}.
It (CSD) decays as a power law with an exponential cutoff in the {\emph{clustered phase}} and tends to a power law as the system approaches the transition to the 
{\emph{phase separated state}}.
We draw a phase diagram in the plane
of size ratio $s$ and volume fraction of passive particles $\phi_p$. 
Our study explores a wide range in the parameter space of ($\phi_p, \phi_a, s$).  

This paper is organized as follows: in Sec.\,\ref{section:model}, we introduce our model of an athermal binary mixture 
of active and passive particles.
The structure and effective interactions in our system are 
presented in Sec.\,\ref{section:struct}. 
We present our results on
cluster size distributions in Sec. 4 before the final section, Sec. 5, where we discuss our results and conclusions.

\section{The model: binary mixture of active and passive discs}
\label{section:model}
Our system consists of a binary mixture of $N_{1}$ small active particles of radius $a_{1}$ and $N_{2}$ big passive 
particles of radius $a_{2}$ $(a_2>a_1)$ moving on a two dimensional frictional substrate. 
Each active 
particle has a self propulsion speed $v$ and its orientation is represented by a unit vector 
$\hat{\boldsymbol{\nu}}_{i}=(\cos\,\theta_{i}, \sin\,\theta_{i})$ where $\theta_i$
is the angle that its velocity makes with respect to some reference direction. The motion of active particles is governed by the following Langevin equations:
\begin{equation}
\partial_{t}{\mathbf{r}_{i}} = v\hat{\boldsymbol{\nu}}_{i}+\mu_{1}\sum_{j\neq i}{\mathbf{F}}_{ij}\,,\label{eom_active}
\end{equation}
\begin{equation}
 \partial_{t}\theta_{i}=\eta^{r}_{i}(t)\,.
\end{equation}
Here $<\eta_{i}^{r}(t)\eta_{j}^{r}(t')>\,=2\nu_{r}\delta_{ij}\delta(t-t')$ where $\nu_{r}$ is the rotational diffusion constant 
of active particles ~\cite{Fily1}\,. $\mu_1$ is the mobility and ${\mathbf{F}}_{ij}$ the force acting on particle $i$ due to particle $j$. 
$\nu_{r}^{-1}$ is the time scale over which the orientation of an active particle changes. Hence, $l_p = v \nu_{r}^{-1}$, the persistence length or run length, is the typical distance travelled by an active particle before it changes direction. In our study, $l_p=20\,a_{1}$\,.  
\begin{figure}[!h]
\begin{center}
{
\includegraphics[scale=0.3,keepaspectratio=true]{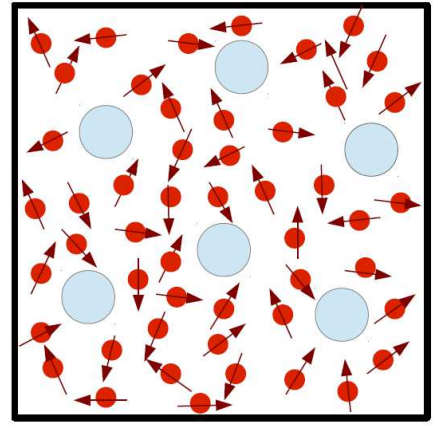}
}
\caption{Schematic diagram of our model: binary mixture of athermal active and passive particles. 
Small particles (with arrow) denote active particles and big particles denote passive particles.}
\label{fig:model}
\end{center}
\end{figure}

The equation of motion for passive particles is 
\begin{equation}
\partial_{t}{\mathbf{r}_{i}} = \mu_{2}\sum_{j\neq i}{\mathbf{F}}_{ij}\,,\label{eom_passive}
\end{equation}
where $\mu_{2}$ is the mobility of passive particles. 
There is no translational noise in Eqs.(\ref{eom_active},\ref{eom_passive}) and hence the particles are athermal. We choose the mobility of both 
species to be the same {\it i.e.}, $\mu_1=\mu_2$\,. 

Particles interact through short ranged soft repulsive 
forces ${\bf{F}}_{ij}=F_{ij}\hat{{\bf{r}}}_{ij}$ where  
$F_{ij}=k(a_{i}+a_{j}-r_{ij})$ if $r_{ij}\le a_{i}+a_{j}$ and $F_{ij}=0$ otherwise; 
$r_{ij}=|{\bf{r}}_{i}-{\bf{r}}_{j}|$ and $k$ is a constant. $(\mu_1 k)^{-1}$ defines
the elastic timescale.

We simulate the system in a square box of size 
$140\,a_1 \times 140\,a_1$ with periodic boundary conditions varying the volume fractions 
$\phi_{a}=N_{1}\pi a_{1}^{2}/L^{2}$ and $\phi_{p}=N_{2}\pi a_{2}^{2}/L^{2}$ 
of active and passive particles respectively. 
We start with a random homogeneous distribution of active and passive particles in the box
and with random directions for the velocity of active particles. Equations (\ref{eom_active}-\ref{eom_passive}) are
updated for all particles and one simulation step is counted after a single update for all the
particles.
The system is defined by the volume fractions $\phi_{a}$ and $\phi_{p}$ of the active and passive particles respectively, the activity $v$ 
of active particles and the size ratio $(\,s=a_2/a_1\,)$ defined as the ratio of the radius of a passive particle to the radius of an active particle. 
We scale the activity by $a_1 \mu_1 k$ to make it dimensionless. The scaled activity is denoted by $v_0\,$. 
We compute the radial distribution functions (RDFs) between pairs of passive particles, and active 
particles to characterise the structure of the mixture and effective interactions between particles. Pure active systems have been shown to form clusters and phase separate 
on increasing $\phi_{a}$ and $v_{0}$\,. To estimate the clustering of both particle species we calculate the cluster size distributions (CSD). 
All data are recorded in the steady state. 

\section{Structure and effective interactions} 
\label{section:struct}
The structure of our binary mixture is given by the distribution of both species
of particles in the mixture.  We identify 3 different phases that characterise the
distribution of passive particles in the mixture --
(i) The homogeneous (mixed) phase - here there is no clustering of passive particles and they are found to be homogeneously distributed. 
The CSD in this phase is exponentially decaying. 
(ii) The clustered phase - here the passive particles form clusters both big and small. 
The CSD decays as a power law with an exponential cutoff indicating that large clusters
though occasionally present are exponentially less probable than small clusters.
(iii) The phase separated phase - here the passive particles segregate into a separate phase. The CSD in this phase is a power law with a peak at large numbers 
indicating the presence of a large cluster or aggregate.    
\begin{figure}[t]
\begin{center}
{
\includegraphics[scale=0.28,keepaspectratio=true]{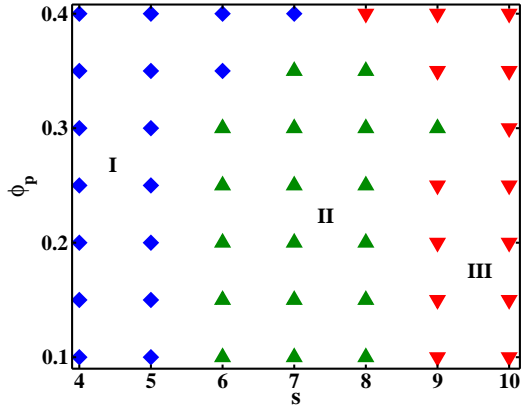}
}
\caption{Phase diagram in $\phi_p$\,-\,$s$ plane. Volume fraction of active particles kept fixed at $\phi_a=0.3$\,. 
Three different phases -- Phase I - homogeneous (mixed) phase, Phase II - clustered phase and 
Phase III - phase separated state, are shown.}
\label{fig:phase}
\end{center}
\end{figure}
Plotted in Fig.~\ref{fig:phase} is a phase diagram in  the $\phi_p$\,-\,$s\,$ plane. In the {\emph {homogeneous or mixed phase}} there is a little or 
no clustering of the passive particles. If atleast $10\%$ of the total number of particles 
are part of clusters, then the system is said to be in the {\emph {clustered phase}}.
 If $40\%$ or more of the particles are in one cluster then the system 
is considered {\emph {phase separated}}. Typical snapshots of the three different phases are shown in 
Fig.~\ref{fig:rdf_size}(b), \ref{fig:rdf_size}(c) and \ref{fig:rdf_size}(d) respectively.
The homogeneous phase extends to larger $s$ for a given $\phi_p$.
The clustered phase occurs at intermediate $s$ and small $\phi_p$. The phase separated phase occurs for large $s$.

To get a qualitative and quantitative picture of  
interparticle interactions and the structure of the binary mixture, we compute the pair distribution functions between pairs of particles. 
The pair distribution function $g({\bf{r}})$ is a measure of the probability of finding a particle at ${\bf{r}}_2$ given a 
particle at ${\bf{r}}_1$\,; ${\bf{r}}=({\bf{r}}_1-{\bf{r}}_2)$. 
For an isotropic system 
$g({\bf{r}})\,\rightarrow\,g(r)$\, and is called radial distribution function (RDF) with $r=|{\bf{r}}|$ ~\cite{chaikin}\,. 
In two dimensions $<n>g({\bf{r}})d^{2}{\bf{r}}$ 
gives the number of particles in $d^{2}{\bf{r}}$\,. 

The structure of the mixture in the steady state and effective interactions are 
characterised by the  radial distribution functions: $g_{pp}(r)$\,, $g_{aa}(r)$ 
of pairs of passive and pairs of active particles respectively 
and  $g_{ap}(r)$ of pairs of active and passive particles. 
We discuss, in the following sections, the features of these  distribution functions 
as a function 
of the various parameters characterising the system -- $\phi_a\,$, the volume fraction of small active particles, $\phi_p\,$, the volume fraction of big 
passive particles and $s$, the size ratio of the particles.
\subsection{Variation with concentration of passive particles}
\begin{figure}[!h]
\begin{center}
{
\includegraphics[scale=0.4,keepaspectratio=true]{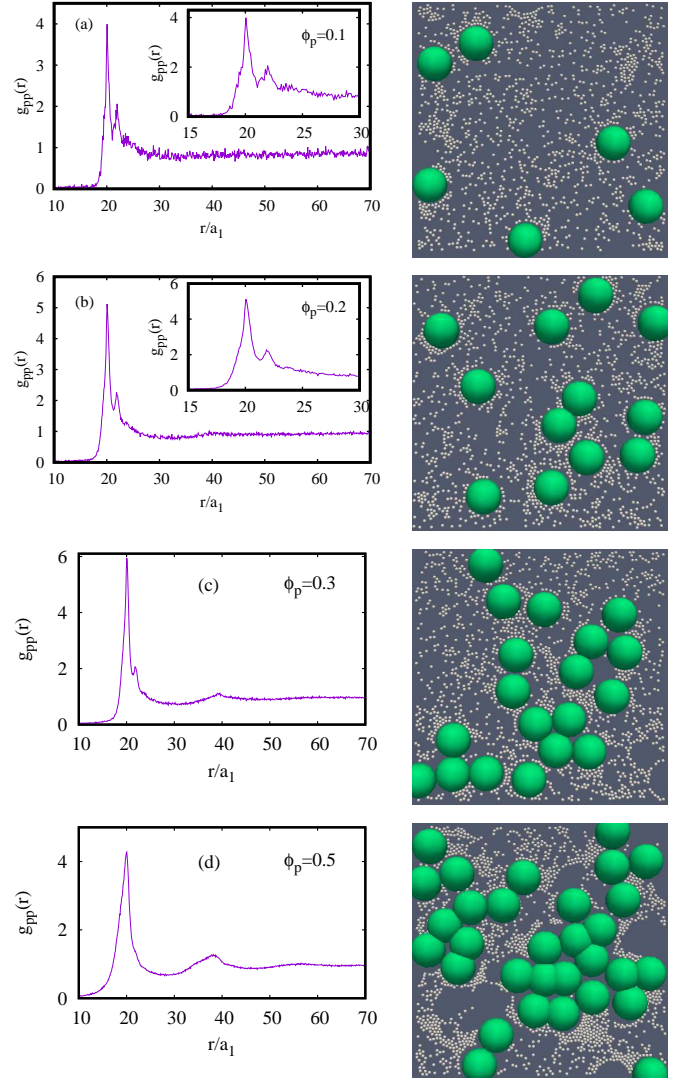}
}
\caption{$g_{pp}(r)$ for different $\phi_p$ at fixed $\phi_a=0.2$ for size ratio $s=10$ and activity $v_0=0.125$\,. Corresponding snapshots are shown in right column. }
\label{fig:rdf_fb}
\end{center}
\end{figure}
\begin{figure}[!h]
\begin{center}
{
\includegraphics[scale=0.6,keepaspectratio=true]{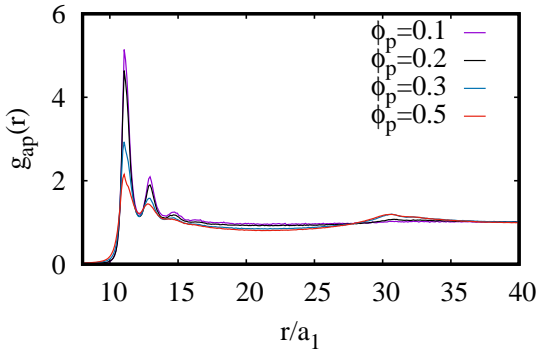}
}
\caption{$g_{ap}(r)$ for different $\phi_p$ at fixed $\phi_a=0.2$ for size ratio $s=10$\,. }
\label{fig:gsb_fb}
\end{center}
\end{figure}

We begin by studying the effect of increasing the concentration of passive particles in the binary mixture for fixed size ratio and concentration of small 
particles. Plotted 
in Fig.\,~\ref{fig:rdf_fb}\,(left column) is the radial distribution function of passive particles for different $\phi_{p}$ at fixed 
size ratio $s=10$\,, activity $v_0=0.125$ and $\phi_{a}=0.2$\,.
 We notice the following main features in $g_{pp}(r)$\,:
(i) for all $\phi_a$, the first peak in $g_{pp}(r)$ occurs at $r=2a_{2}$ corresponding to the diameter of the passive particle. 
(ii) For small $\phi_p$, the second peak occurs at $2(a_1+a_2)$ indicating a layer of active particles around the passive particles. Active particles 
have a tendency to aggregate. Passive particles, being less mobile, seed 
this motility induced aggregation of active particles. This results in layer(s) of
active particles surrounding the passive particle which causes the second peak to appear at $r=2(a_1+a_2)\,$.    
There is no clustering of passive particles at such low volume fractions.
(iii) At sufficiently large $\phi_p$, a peak appears at $r \le 4a_2$ indicating the formation of clusters of passive particles. 
At larger $\phi_p$, this peak becomes more pronounced, broadens and shifts to 
$r < 4a_2$ indicating close packing of particles in the cluster and 
the formation of large clusters. In this configuration the peak at 
$r=2(a_1+a_2)\,$ is not seen implying that the passive particles are mostly together
 in clusters and the interface with active particles is minimized. 
Snapshots of the system corresponding to the RDFs are shown in right column of Fig.\,~\ref{fig:rdf_fb}\,. Passive particles begin to form cluster for $\phi_p > 0.25$\,. 
This clustering can be attributed to an effective attraction between them due to the surrounding active particles. 
In all cases, the effective interaction between the small active and big passive
particles is attractive (see $g_{ap}$ in Fig.\ref{fig:gsb_fb}). Since there exists a fixed concentration of active particles, not all passive particles 
can be surrounded by active particles as their concentration is increased. They  hence cluster and the clustering increases with increasing concentration of big particles.

The effective interaction between two passive particles must also
depend on the concentration of small particles and the size ratio of particles.
To understand this dependence, we first analyse the behaviour of the mixture at different $\phi_a$ keeping $\phi_p$, $s$ and the activity fixed. 
\subsection{Variation with concentration of active particles}
\begin{figure}[!h]
\begin{center}
{
\includegraphics[scale=0.4,keepaspectratio=true]{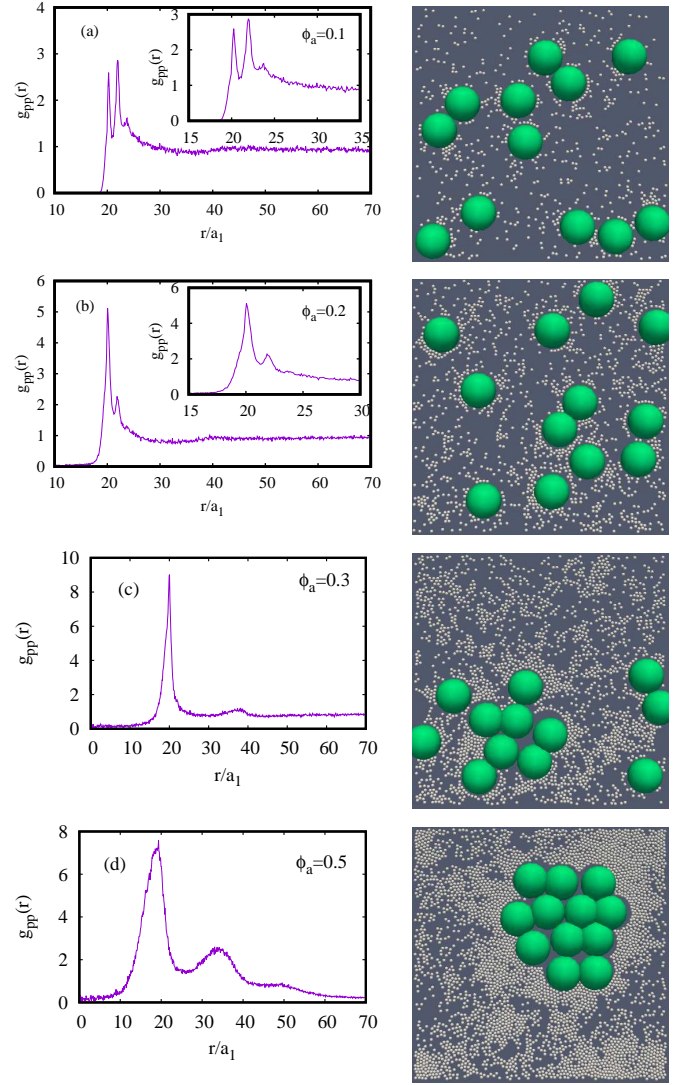}
}
\caption{$g_{pp}(r)$ for different $\phi_a$ at fixed $\phi_p=0.2$ for size ratio $s=10$ and activity $v_0=0.125$\,. 
 Corresponding snapshots are shown in right column.}
\label{fig:rdf_fs}
\end{center}
\end{figure}

Plotted in Fig.\,~\ref{fig:rdf_fs}\,(a)-(d) is the RDF $g_{pp}(r)$ for different $\phi_a$ at fixed $\phi_p=0.2$ and activity $v_0=0.125$\,. 
For small $\phi_a < 0.2\,$, the first peak occurs at $2a_2$ and a second 
larger peak occurs at $2(a_1+a_2)$ indicating 
the presence of a layer of small particles around the passive particles. A third
much smaller peak at $2(a_2+2a_1)$ indicates a second layer
of active particles around some of the passive particles. $g_{pp}(r)$ decays to $1$ at large distances. 
The tendency of active particles to aggregate in regions of low motility (motility induced aggregation) leads to their increased presence around the passive particles. 
But as we increase $\phi_a$, the peak at $2(a_2+a_1)$ diminishes and
$g_{pp}(r)$, starts to show signatures of a peak at $4a_2$.
This is indicative of the beginning of clustering of big particles and the presence of fewer isolated passive particles. 
The concentration of active particles is now sufficient to lead to a significant attractive interaction between the passive particles 
and cause clustering. For larger $\phi_a=0.3$, the peak at $2(a_1+a_2)$ disappears and the second peak is located close to but 
less than $r=4a_2$. This signals the formation of dense clusters of passive particles and the absence of isolated ones (as seen in the corresponding snapshot). At very 
large $\phi_a$ ($\ge 0.5$), the second peak is very clearly located at $2\sqrt{3}a_2$ indicating hexagonal close packed clusters of passive particles
which is also clear from the corresponding snapshot (see Fig.\,~\ref{fig:rdf_fs}(d) right column). Also
$g_{pp}(r)$ decreases below $1$ for large $r$. This is an indication of complete phase separation. 

\begin{figure}
\begin{center}
{
\includegraphics[scale=0.7,keepaspectratio=true]{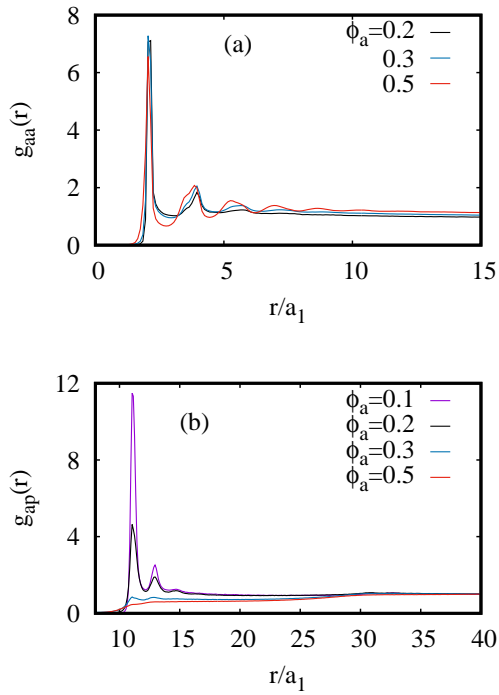}
}
\caption{(a) $g_{aa}(r)$ for different concentrations of active particles. (b) $g_{ap}(r)$ for different concentrations of active particles 
at fixed $\phi_p=0.2$ for size ratio $s=10$\,. }
\label{fig:gsb_gss_fs}
\end{center}
\end{figure}

Shown in Fig.\ref{fig:gsb_gss_fs}(a), are the RDFs between pairs of active particles
$g_{aa}(r)$ for different $\phi_a$. 
The first peak occurs at $2a_{1}$ and second peak at $4a_{1}$\,. 
The second peak has a structure. A small peak appears at $r=2\sqrt{3}a_{1}$\,. This small peak is more 
prominent for large $\phi_{a}$ and indicates hexagonal closed packed (HCP) structures in the cluster of small particles. 
Broadening of the peaks indicates squeezing of small particles for large $\phi_{a}$\,. Fig.\ref{fig:gsb_gss_fs}(b) shows the corresponding $g_{ap}(r)$. 
As the concentration $\phi_a$ increases the attractive interaction at short range between
active and passive particles diminishes. At $\phi_a=0.5$, it is completely absent
and the interaction at $(a_1+a_2)$ is repulsive. This signals complete phase separation. 
This is unlike the formation of large clusters in Fig.\,~\ref{fig:rdf_fb}(d)\,.
\subsection{Variation with size of passive particles}
\begin{figure}[!h]
\begin{center}
{
\includegraphics[scale=0.4,keepaspectratio=true]{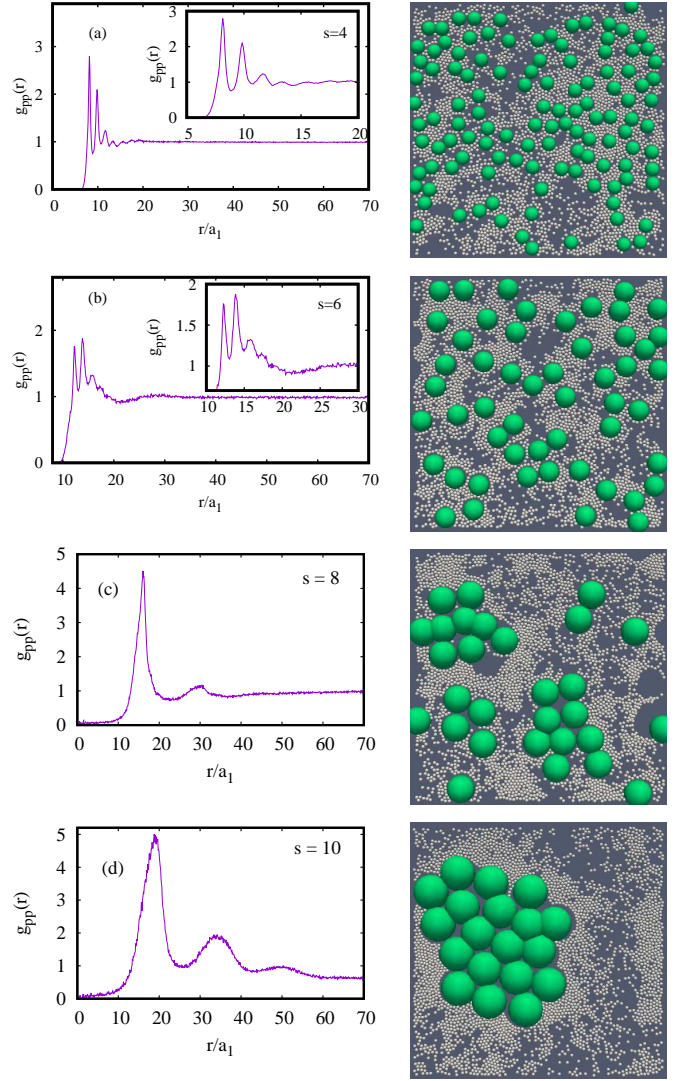}
}
\caption{$g_{pp}(r)$ for different size ratios at fixed $\phi_p=0.3$\,, $\phi_a=0.4$ and activity $v_0=0.125$\,. Corresponding snapshots are shown in right column. 
}
\label{fig:rdf_size}
\end{center}
\end{figure}

\begin{figure}[!h]
\begin{center}
{
\includegraphics[scale=0.6,keepaspectratio=true]{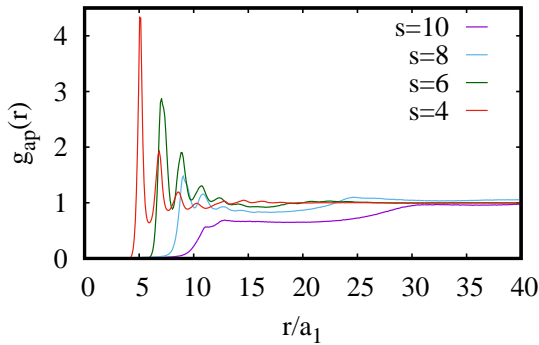}
}
\caption{$g_{ap}(r)$ for different size ratios at fixed volume fractions $\phi_a=0.4$ and $\phi_p=0.3$\,.}
\label{fig:gsb_size}
\end{center}
\end{figure}

The force on a passive particle depends crucially on the size ratio 
of particles in our system. We have chosen the mobility of a passive particle to
be a constant independent of its size. The velocity imparted to a passive
particle depends on its size because of the number of active particles that can strike it over a short duration of time. 
To understand better the role of size on the effective interaction between passive particles in our system, we vary the 
size of the passive particle keeping all other parameters fixed in our system.  
The pair distribution functions $g_{pp}(r)$ for different size ratios $s=4,\, 6,\, 8$ and $10$ are plotted in Fig.~\ref{fig:rdf_size}\,(left column) for 
fixed $\phi_p=0.3$ and $\phi_a=0.4$ and activity $v_0=0.125$ of active particles\,. 
The size of passive particles is changed keeping the size of active particles fixed.
For small $s \le 4$, we find pairs and groups of 3 passive particles clustered together in the surrounding medium of active particles. 
The first peak in $g_{pp}(r)$ at $r=2 a_2$ is the dominant peak and subsequent peaks (decreasing in magnitude) are observed at a 
separation of $2 a_1$ from this one. These are indicative of layers of small particles around some of the passive particles. 
For slightly larger $s$, the passive particles are isolated (from each other) and surrounded by layers of active particles. The 
peak at $r=2 (a_2+a_1)$ is now the dominant one indicating that a layer of active particles around a passive particle is 
the most favoured configuration (see Fig.\ref{fig:rdf_size}(b))\,. Higher order peaks are observed at a separation of $2 a_1$, as for smaller $s$, but they are more 
pronounced here ~\cite{Mani}\,. In both the cases, the dynamics and structure of passive particles is dictated by the active 
particles and predominantly by their tendency to aggregate. Here again, passive particles
act as regions of low mobility and facilitate motility induced aggregation of active particles.  
As the asymmetry in size increases, the second peak appears close to 
$r=4 a_2$. This is indicative of clustering of passive particles. For large
size ratios $s \ge 7$, the effective attractive interaction between passive 
particles induced by the active particles is significant. 
For size ratio $s=10$, a strong clustering of passive particles leads to 
complete phase separation. The first peak occurs at 
$2 a_{2}$, the diameter of the passive particle, and the second peak appears at $2\sqrt{3} a_{2}$\,. This indicates that the particles are hexagonally 
close packed inside the cluster. The corresponding snap shot in Fig.\ref{fig:rdf_size}(d) also reveals the same.   
$g_{pp}(r)$ tends to a value $< 1.0$ at large $r$ indicating complete phase separation.
Fig.~\ref{fig:rdf_size}\, (right column) shows snapshots for different size ratios at fixed $\phi_p=0.3$ and $\phi_a=0.4$ and activity $v_0=0.125$ of active particles\,.
On examining $g_{ap}$ plotted in Fig.~\ref{fig:gsb_size}, it is seen the interaction is attractive 
at short range for small $s (\le 8) $ and repulsive at large $s (=10)$.
The structure of our binary mixture can be largely understood from the tendency of active particles to phase separate into a 
dense solid-like phase and a dilute gas-like phase.
This phenomenon, known as motility induced phase separation,
results from the growth of a small fluctuation in the local density of active particles 
~\cite{Fily1,Fily2,Simha,Sriram1,Redner,Cates1,Tailleur1,Gonnella1,Gonnella2,Berthier,Cates2}\,. Particles in this dense region 
are slower than those elsewhere because of their 
increased density (or enhanced crowding). Particles in the vicinity of this region slow down as they approach it, again because 
of crowding, and become a part of it on slowing down sufficiently. This positive feedback causes the fluctuation to grow and result 
in a dense macroscopic aggregate of active particles. Particles far away from this region remain in a dilute phase so the result is 
a phase separated system. 
In our binary mixture, the dynamics of passive particles comes largely from their 
interactions with active particles. For a large range of parameters, passive particles act as regions of low motility and seed MIPS, 
just as a small region
of increased density does in pure active systems. Active particles in the mixture
are hence attracted to passive particles and accumulate around them. This effective attractive interaction can be deduced from the 
$g_{ap}$ plotted in Fig.~\ref{fig:gsb_size}.  The majority of passive particles are hence embedded inside aggregates of active particles. 
Depending on the concentration of active particles and size ratio, single or small clusters of passive particles are formed. The cluster 
size decreases on increasing the size of passive particles (for a fixed concentration of active particles) so that the (active-passive particle) 
interface length is more or less constant. $g_{pp}$ has peaks at 
$2(a_1+a_2)$ and successively at a separation of $2 a_1$. An effective attractive interaction between passive particles exists.   

\begin{figure}[!h]
\begin{center}
{
\includegraphics[scale=0.6,keepaspectratio=true]{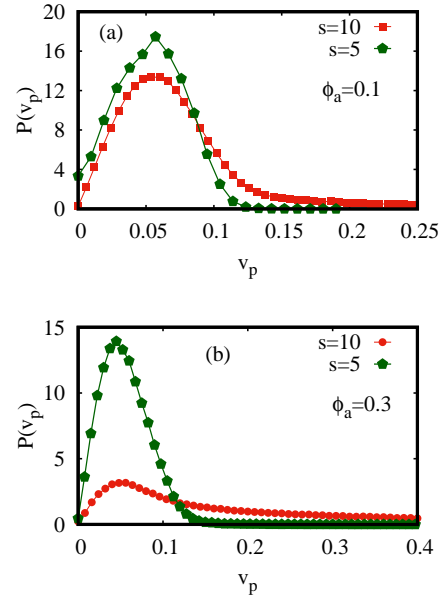}
}
\caption{Velocity distribution of passive particles for $s=5$ and $s=10$ for $\phi_a=0.1$ and $0.3$\,. 
For all these set of data activity is kept constant at $v_0=0.125$\,.}
\label{fig:vel_dist}
\end{center}
\end{figure}

For large $\phi_a$ and sufficiently large size ratio $s$, the passive particles are faster as they receive more kicks from the active particles 
and hence acquire a larger velocity. Plotted in Fig.~\ref{fig:vel_dist}\,.
are the velocity distributions of passive particles for 2 size ratios. At low $\phi_a$,
these distributions do not vary significantly as the number of active particles coming close to the passive particle over a short duration is small (see Fig.~\ref{fig:vel_dist}(a))\,. 
At large $\phi_a$, big passive particles are much faster than the small ones (see Fig.~\ref{fig:vel_dist}(b))\,. 
Passive particles no longer seed MIPS as they are not slow and in fact, their effective interaction 
with active particles is now repulsive at short range. This is seen in the plots for $g_{ap}$ in Fig.~\ref{fig:gsb_size}. In this parameter range, 
the passive particles phase separate
or form large clusters to stay away from active particles. This kind of behaviour can be seen in Figs.~\ref{fig:rdf_fs}(d)\,,~\ref{fig:rdf_size}(d)\,.
  
\section{Cluster size distributions}
\label{section:csd}

\begin{figure}[!h]
\begin{center}
{
\includegraphics[scale=0.47,keepaspectratio=true]{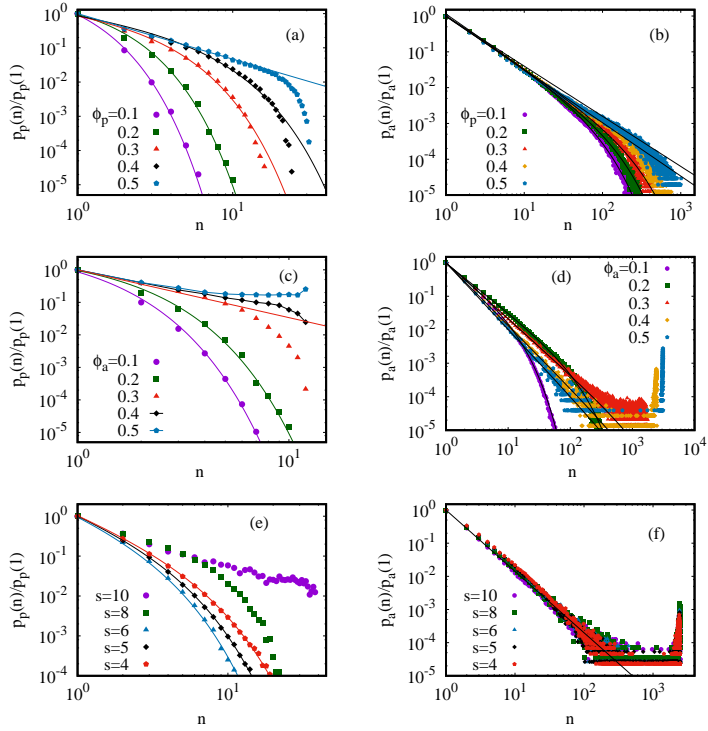}
}
\caption{Cluster size distributions of passive and active particles for different sets of parameters. 
(a) CSD of passive particles for different $\phi_p$ at constant $\phi_a=0.2$ for $s=10$\,. (b) Corresponding CSD of active particles.  
(c) CSD of passive particles for different $\phi_a$ at constant $\phi_p=0.2$ for $s=10$\,. (d) Corresponding CSD of active particles. 
(e) CSD of passive particles for different size ratio at fixed $\phi_p=0.3$ and $\phi_a=0.4$\,. (f) Corresponding CSD of active particles.  
Activity is kept constant at $v_0=0.125$ for all these sets of parameters.}
\label{fig:csd1}
\end{center}
\end{figure}

The cluster size distributions (CSD) of both active and passive particles for different volume fractions and size ratios are plotted in Figs.~\ref{fig:csd1}. 
Plots in the left column of Fig.~\ref{fig:csd1}
are CSDs of passive particles while those in the right column are CSDs for active particles in the binary mixture. 
We begin by analysing the effect of varying the volume fraction of passive particles, $\phi_p$ in the mixture. Fig.~\ref{fig:csd1}(a) 
shows the CSD of passive particles $p_{p}(n)$ for different $\phi_p$ at fixed activity $v_0=0.125$\,, size ratio 
$s=10$ and $\phi_a=0.2$\,. For small $\phi_p < 0.3$\,, 
$p_p(n)/p_p(1)$ has the exponential form $exp(-n/n_0)$ as observed in the thermal case.
There is no clustering at these volume fractions. For intermediate 
volume fractions $0.3 < \phi_p < 0.5$\,, $p_p(n)$, shows a power law decay with exponential cutoff at large 
$n$, {\it i.e.}, $p_p(n)/p_p(1) \simeq \frac{1}{n^{\alpha}}\exp(-n/n_0)$
for $\phi_p=0.3$, $0.4$\,.  As we further
increase $\phi_p > 0.4$\,, the distribution is predominantly a power law with a 
sudden decay for large $n$. Here the average cluster size is very large (of the
order of the system size). 

Correspondingly, the CSDs of active particles fits to the form  $p_a(n)/p_a(1) \simeq \frac{1}{n^{\alpha}}\exp(-n/n_0)$, a power law at 
small $n$ followed by an exponential decay at large $n$, for volume fractions $\phi_p \le 0.3$\,. For larger $\phi_p$, the distribution is 
predominantly a power law. Shown in Fig.\,~\ref{fig:csd1}\,(b) is a power law fit $1/n^{1.49}$ for $\phi_p=0.4$\,. 

Next, we study the effect of the concentration of active particles on the clustering of big passive particles.
In Fig.\,~\ref{fig:csd1}\,(c) we plot $p_p(n)/p_p(1)$ for fixed $\phi_p=0.2$ and for different $\phi_a$ (values indicated in the plot). For very small
$\phi_a \le 0.2$, $p_p(n)/p_p(1)$ decays exponentially with $n$, which represents 
the homogeneous state or no clustering. For larger $\phi_a=0.3$\,, 
$p_p(n)/p_p(1) \simeq \frac{1}{n^{\alpha}}$\,, with power $\alpha=1.47$\,. At very large $n$, the distribution drops suddenly. The CSDs for 
larger $\phi_a$ cannot be fitted to any of these forms.

The corresponding CSDs for active particles are plotted in 
Fig.\,~\ref{fig:csd1}\,(d). For small $\phi_a$, the distribution is a power law for small $n$ followed by an exponential drop at large 
$n$\,.\,$p_a(n)/p_a(1) \simeq \frac{1}{n^{\alpha}}\exp(-n/n_0)$ as before with the 
power $\alpha = 1.45, 1.47$ respectively for $\phi_a = 0.1\,,0.2$\,. 
For large volume frcations, the CSDs shows a bimodal distribution with a clean
power law for range of n with a large peak at large n indicating the formation of large
aggregates. This indicates a phase separated state.
The distribution shifts to the left on increasing $\phi_a$ while the height of the peak at large $n$ increases. This implies that small clusters 
become less probable as the formation of one
large aggregate is favoured more and more as $\phi_a$ increases. The power law exponent we find is close to 1.5, which is similar to the exponent find
in the study of Peruani et al. for system interacting with short range interaction. Also
bimodal nature of CSDs agrees well with the predictions of Peruani et al. {\it et al}~\cite{Peruani1}\,.

\begin{figure}[!h]
\begin{center}
{
\includegraphics[scale=0.6,keepaspectratio=true]{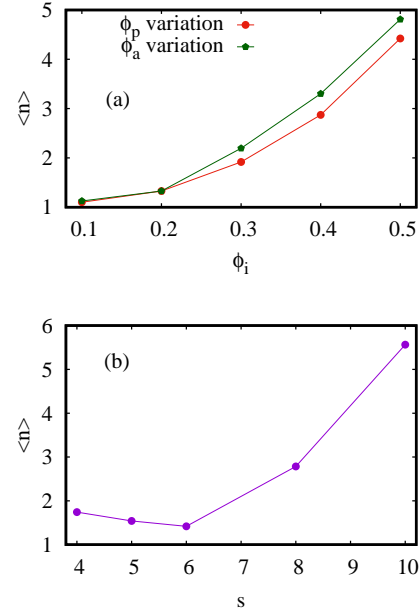}
}
\caption{(a) Average cluster size of passive particles for different $\phi_a$ and $\phi_p$ at fixed size ratio $s=10$\,. (b) Average cluster size 
of passive particles for different size ratios at fixed volume fractions $\phi_p=0.3$ and $\phi_a=0.4$\,.}
\label{fig:av_cl}
\end{center}
\end{figure}

Fig.\,~\ref{fig:csd1}(e) depicts the $p_p(n)/p_p(1)$ for fixed $\phi_a=0.4$, $\phi_p=0.3$ for different sizes of passive particles, 
keeping the size of active particles fixed at $a_1=0.1$. As the size ratio increases from $s=4$ to $s=6$\,, clustering of passive particles 
decreases. The mean cluster size drops $s=4$ to $s=6$\, (see Fig.\,~\ref{fig:av_cl}(b)). As before the distributions fit well to a power law with an 
exponential cut off, with the power $\approx 1.5$\,. 
For $s=8,10$, the CSDs do not fit to any known functional forms, however it is clear from the distributions
that the  clustering increases with increasing $s$.
The CSDs for active particles for the same parameters, 
shown in  Fig.\,~\ref{fig:csd1}(f), nearly obey a power law but with peaks at large $n$ indicating that they are phase separated. This is also 
clear from the snapshots
in Fig.\,~\ref{fig:gsb_size} where dense aggregates of small particles can be seen. 

We summarise our results on the CSDs of active and passive particles. At very
low concentrations or volume fractions, the CSD resembles that at equilibrium and has an exponential form $exp(-n/n_0)$\,. 
At intermediate concentrations, the distributions are well
described by a power law with an exponential cutoff $n_0$\,,
$\frac{1}{n^{\alpha}}\exp(-n/n_0)$\,. As the system approaches the
transition to phase separate, the CSD is very nearly a power law $1/n^{\alpha}$ but with a sudden
drop at large cluster sizes.  The power $\alpha \rightarrow 1.5$ as the phase transition is approached. 
Power law decay of CSD is a generic feature
as transition to phase separated state is approached. This has previously been 
obtained in numerical simulations of active systems \cite{Peruani1}\,. A simple kinetic model to describe clustering 
in active or SPP systems is presented in the work of Peruani ${\textit {et al.}}$\cite{Peruani1}\,.
The model incorporates the description of clustering in terms of coagulation and fragmentation of 
clusters conserving only the total number of particles. In terms of the CSD two distinct phases are 
obtained (i) an individual phase, where the CSD is
dominated by an exponential tail that defines a characteristic cluster size, and (ii)
a collective phase characterized by the presence of a non-monotonic CSD with
a local maximum at large cluster sizes. A critical point separates the two phases where 
the CSD is a power law $p(n) \sim n^{-\gamma}$\,, with the exponent $\gamma$ lying in the range 
$(0.8, 1.5)$ for different parameters that define the details of the system.

In the phase separated state, the CSD is very
nearly  a power law but with a peak at large $n$ indicating the formation of large aggregates. Small clusters become less probable and the 
formation of one
large aggregate is favoured more and more as one moves deeper into the phase separated regime as in Peruani {\it et al}\cite{Peruani1}\,. The mean cluster size increases 
monotonically with the concentration of particles $\phi_p,\,\phi_a$ (see Fig.\,~\ref{fig:av_cl}(a)\,)\,. It decreases, initially for small $s\,$,
and then increases with $s$ for large $s \,(\ge 7)$ as shown in Fig.\,~\ref{fig:av_cl}(b)\,. 
%
%
\section{Conclusions}
\label{section:conclusions}
We have studied the dynamics of binary mixtures of small active and big passive athermal particles, interacting via soft repulsive forces. 
The motility of small active particles induces an effective attraction between
big passive particles. 
This attraction leads to true phase separation for large size ratios and volume fraction of active particles. For small volume fractions of active particles and upto 
intermediate size ratios, passive particles being slower than active particles seed MIPS.
 The effective interaction between active and passive particles is attractive and the cluster size decreases with increasing 
 size ratio. For larger size ratios and concentration of active particles, passive particles are fast and the interaction between active and passive particles 
 can become repulsive at short range. The mean cluster size now increases with increasing size ratio. Three distinct phases of passive particles can be identified based 
 on their spatial distribution. At
low volume fractions and small size ratios, the CSD of passive particles resembles that at equilibrium and has an exponential form $exp(-n/n_0)$\,. 
This is referred to as the {\emph {homogeneous phase}}.
At intermediate size ratios, the CSD of passive particles decays as a power law with an exponential cutoff (in the {\emph {clustered phase}}) and tends to a pure power law as 
the system approaches the transition to the {\emph {phase separated}} state for large size ratios. 
While there have been many studies on mixtures of active passive particles, they are mostly limited to either very few passive
particles in an active medium or to  active/passive particles that are non-spherical: like the recent study of passive plates in active medium \cite{Ni}, 
self-propelled active rods interacting with passive particles \cite{Yen}\,.
Our study explores the phase behaviour when both types of particles are comparable in volume fraction and for a wide range of parameters defining a binary mixture.

\section*{Acknowledgement}
We thank HPCE, IIT Madras for providing computing facilities. SM thanks S. N. Bose National Center for Basic Sciences, Kolkata for its kind hospitality.

\bibliography{ap} 
\bibliographystyle{rsc} 

\end{document}